\documentclass{article}

\usepackage{amsmath}
\usepackage{cite}
\newcommand{\Mu}{\mathrm{M}}
\newcommand{\h}{\mu}
\newcommand{\st}{\mathrm{st}}
\newcommand{\sgn}{\mathrm{sgn}}
\newcommand{\fb}{\mathbf{f}}
\begin{document}
\begin{titlepage}
\begin{center}
{\large \textbf{One-dimensional kinetic Ising model with
nonuniform coupling constants}} \vskip 2\baselineskip \centerline
{\sffamily Mohammad Khorrami \footnote{e-mail:mamwad@mailaps.org}
\&
 Amir Aghamohammadi\footnote
{mohamadi@alzahra.ac.ir}}
 \vskip 2\baselineskip
{\it Department of Physics, Alzahra University, Tehran 19384,
IRAN}
\end{center}


\begin{abstract}
\noindent A nonuniform extension of the Glauber model on a
one-dimensional lattice with boundaries is investigated. Based on
detailed balance, reaction rates are proposed for the system. The
static behavior of the system is investigated. It is shown that
there are cases where the system exhibits a static phase
transition, which is a change of behavior of the static profile of
the expectation values of the spins near end points.
\end{abstract}
\end{titlepage}
\newpage
\section{Introduction}
The Glauber dynamics was originally proposed to study the
relaxation of the Ising model near equilibrium states. It is a
simple non-equilibrium model of interacting spins with spin-flip
dynamics. It is also known that there is a relation between the
kinetic Ising model at zero temperature and the diffusion
annihilation model in one dimension. There is an equivalence
between domain walls in the Ising model and particles in the
diffusion annihilation model. Kinetic generalizations of the Ising
model, for example the Glauber model or the Kawasaki model, are
phenomenological models and have been extensively studied
\cite{RG,KK,AF,TV,SSG,GL}. Combination of the Glauber and the
Kawasaki dynamics has been also considered \cite{DRS,GF,AT}. Most
studies are focused on completely uniform lattices with
site-independent rates. Among the simplest generalizations beyond
a completely uniform system is a lattice with alternating rates.
In \cite{SchSch02,SchSch03,MObZ}, the steady state configurational
probabilities of an Ising spin chain driven out of equilibrium by
a coupling to two heat baths has been investigated. An example is
a one-dimensional Ising model on a ring, in which the evolution is
according to a generalization of Glauber rates, such that spins at
even (odd) lattice sites experience a temperature $T_e$ ($T_o$).
In this model the detailed balance is violated. The response
function to an infinitesimal magnetic field for the Ising-Glauber
model with arbitrary exchange couplings has been studied in
\cite{Chatelain}. Other generalizations of the Glauber model
consist of, for example, alternating-isotopic chains and
alternating-bound chains (\cite{GO} for example).

In \cite{MA}, an asymmetric generalization of the zero-temperature
Glauber model on a lattice with boundaries was introduced. There
it was shown that in the thermodynamic limit, when the lattice
becomes infinite, the system shows two kinds of phase transitions.
One of these is a static phase transition, the other a dynamic
one. The static phase transition is controlled by the reaction
rates, and is a discontinuous change of the behavior of the
derivative of the stationary magnetization  at the end points,
with respect to the reaction rates. The dynamic phase transition
is controlled by the spin flip rates of the particles at the end
points, and is a discontinuous change of the relaxation time
towards the stationary configuration. Other phase transitions
induced by boundary conditions have also been studied
(\cite{HS,RIK,AM2} for example). Another generalization of the
Glauber model was introduced in \cite{SAM}. In this
generalization, the processes are the same as those of the
ordinary Glauber model, but the rates depend on three free
parameters, rather than one free parameter in the ordinary Glauber
model. Finally, this model was further generalized to the case
where the number of interacting sites is more than three and the
number of states at each site is more than two. This model too
violates detailed balance.

In the present paper an Ising model on a nonuniform lattice with
boundaries is investigated. Detailed balance is used to propose
reaction rates for the system. Based on this, the evolution of the
expectation values of spins is obtained. The time-independent
solution to this equation is studied. This solution satisfies a
homogeneous difference equation of the second order in the bulk,
the solution to which can be expressed in terms of a transfer
matrix. The reactions at the boundaries impose nonhomogeneous (but
at most linear) boundary conditions on this solution, which could
be used to fix the constants appeared in the static solution.
While it is true that the ordinary Ising model does not exhibit
any phase transition in finite temperature (the expectation values
of the spins vanish if there is no external magnetic field), this
is not necessarily the case for the model studied here. The
expectation values do not vanish as a result of inhomogeneous
boundary conditions. It is shown that in the thermodynamic limit
(when the size of the lattice tends to infinity) different phases
could occur for this system, according to whether the eigenvalues
of the transfer matrix are less than or larger than one. While
this transition (or even the existence of nonzero static
solutions) is invoked by the presence of inhomogeneous boundary
conditions, the detailed form of the boundary conditions affects
only the coefficients of the eigenvectors of the transition matrix
in the static solution. So the detailed form of the boundary
conditions do not affect the static phase portrait of the system.
A closed form is obtained for this transfer matrix, and some
examples are discussed, specially one example in which a phase
transition is seen.

The scheme of the paper is as follows. In section 2, the model is
introduced, the rates are determined using the detailed balance
criterion, and the evolution equation for the spin expectation
values is obtained. In section 3, the time-independent solution is
studied, and the corresponding phase portrait is investigated. In
section 4, some examples are studied in more detail, specially one
example which exhibits a static phase transition. Section 5 is
devoted to the concluding remarks.

\section{One-dimensional Ising model with nonuniform coupling constants}
Consider a one-dimensional lattice with $(L+1)$ sites, labeled
from $0$ to $L$. At each site, there is a spin variable, $s_i$,
which could be $+1$ for spin up ($\uparrow$), or $-1$ for spin
down ($\downarrow$). These spins in the bulk ($s_i$'s with
$0<i<L$) interact according to the Ising Hamiltonian,
\begin{equation}\label{ihg.01}
{\mathcal H} =-\sum_\alpha J_\alpha\,s_{\alpha-\h}\,
s_{\alpha+\h},
\end{equation}
where $J_\alpha$ is the coupling constant in the link $\alpha$,
and
\begin{equation}\label{ihg.02}
\h:=\frac{1}{2}.
\end{equation}
The link $\alpha$ links the sites $\alpha-\h$ and $\alpha+\h$, so
that $\alpha\pm\h$ are integers, and $\alpha$ runs from $\h$ up to
$(L-\h)$.

The usual Glauber model gives the dynamics of the Ising model with
uniform coupling constants ($J_\alpha$ independent of $\alpha$,
and denoted by $J$) such that the rate of a spin flip is
determined through its interaction with its two neighboring sites
and a heat bath at temperature $T$. A spin is flipped with the
following rates.
\begin{align}\label{ihg.03}
\uparrow \; \uparrow \; \uparrow \; \to \; \uparrow \; \downarrow \;
 \uparrow \; \mbox{ and }
\downarrow \; \downarrow \; \downarrow \; \to \; \downarrow \;  \uparrow
\; \downarrow \; & \mbox{\quad with rate\quad}  1-\tanh(2\,K), \nonumber \\
\uparrow \; \downarrow \; \uparrow \; \to \; \uparrow \; \uparrow
\; \uparrow \; \mbox{ and } \downarrow \; \uparrow \; \downarrow \; \to
 \; \downarrow \;  \downarrow \;
\downarrow \; & \mbox{\quad with rate\quad}  1+\tanh(2\,K), \nonumber \\
\uparrow \; \uparrow \; \downarrow \; \rightleftharpoons \;
\uparrow \; \downarrow \; \downarrow \; \mbox{ and } \downarrow \;
\downarrow \; \uparrow \; \rightleftharpoons \; \downarrow \;
\uparrow \; \uparrow \; & \mbox{\quad with rate\quad}  1,
\end{align}
where
\begin{equation}\label{ihg.04}
K:=\frac{J}{k_\mathrm{B}\,T},
\end{equation}
and $k_\mathrm{B}$ is the Boltzmann's constant. As it is seen,
similar to the Ising model, the Glauber model has left-right and
up-down symmetries. The Glauber model has also a particle
reaction-diffusion interpretation. One considers a link with
different spins at its sites (a domain wall) a particle
($\bullet$), and a link with same spins at its sites (no domain
wall) a hole ($\circ$). Then the Glauber model turns into a
reaction-diffusion model:
\begin{align}\label{ihg.05}
\circ\, \circ \to \, \bullet\, \bullet &\mbox{\quad with rate\quad} 1-\tanh(2\,K), \nonumber \\
\bullet\, \bullet \to \, \circ\, \circ &\mbox{\quad with rate\quad} 1+\tanh(2\,K), \nonumber \\
\bullet\, \circ \rightleftharpoons \, \circ\, \bullet &\mbox{\quad
with rate\quad} 1.
\end{align}
Consider the general case where the coupling constant is not
uniform (and the interaction is not necessarily nearest neighbor).
Assuming that in each step only one spin flips, detailed balance
gives
\begin{align}\label{ihg.06}
\frac{\omega(\cdots, s_j,\cdots\to \cdots,-s_j,\cdots)}
{\omega(\cdots, -s_j,\cdots\to \cdots,s_j,\cdots)}&=
\frac{\exp(\cdots+\sum_{i\ne j}K_{i\,j}\,s_i\,(-s_j)+\cdots)}
{\exp(\cdots+\sum_{i\ne j}\,K_{i\,j}\,s_i\,s_j+\cdots)},
\nonumber\\&=\frac{\exp(-h_j\,s_j)}{\exp(h_j\,s_j)},
\end{align}
where $\omega$ is the rate, $K_{i\,j}$ is defined like
(\ref{ihg.04}) but with $J_{i\,j}$ (the coupling between sites $i$
and $j$) instead of $J$, and
\begin{equation}\label{ihg.07}
h_j:=\sum_{i\ne j}K_{i\,j}\,s_i.
\end{equation}
As the value of $s_i$ is either $1$ or $-1$, any function of $s_i$
is at most linear in $s_i$. One then arrives at
\begin{equation}\label{ihg.08}
\exp(h_j\,s_j)=\cosh h_j+s_j\,\sinh h_j.
\end{equation}
Using these, (\ref{ihg.06}) gives
\begin{equation}\label{ihg.09}
\frac{\omega(\cdots, s_j,\cdots\to
\cdots,-s_j,\cdots)}{\omega(\cdots, -s_j,\cdots\to
\cdots,s_j,\cdots)}=\frac{1-s_j\,\tanh h_j}{1+s_j\,\tanh h_j},
\end{equation}
or
\begin{equation}\label{ihg.10}
\omega(\cdots, s_j,\cdots\to
\cdots,-s_j,\cdots)=\Gamma_j\,(1-s_j\,\tanh h_j),
\end{equation}
where $\Gamma_j$'s are constants. In the simple case of nearest
neighbor interaction, one has
\begin{equation}\label{ihg.11}
J_{i\,j}=J_{i+\h}\,\delta_{i, j-1}+J_{i-\h}\,\delta_{i, j+1},
\end{equation}
so that (\ref{ihg.10}) becomes
\begin{equation}\label{ihg.12}
\omega(\cdots, s_j,\cdots\to \cdots,-s_j,\cdots)=\Gamma_j\,
[1-s_j\,\tanh(K_{j-\h}\,s_{j-1}+K_{j+\h}\,s_{j+1})],\qquad
\end{equation}
So the spin at the site $j$ flips according to following
interactions with the indicated rates,
\begin{align}\label{ihg.13}
\uparrow \; \uparrow \; \uparrow \; \to \; \uparrow \;
\downarrow\; \uparrow \mbox{\quad and\quad} \downarrow \;
\downarrow \; \downarrow\; \to \; \downarrow \; \uparrow \;
\downarrow &\mbox{\quad with rate\quad} 1-\tanh(K_{j-\h}+K_{j+\h}), \nonumber \\
\uparrow \; \downarrow \; \uparrow \; \to \; \uparrow \;
\uparrow\; \uparrow \mbox{\quad and\quad} \downarrow \; \uparrow
\; \downarrow\; \to \; \downarrow \; \downarrow \;
\downarrow &\mbox{\quad with rate\quad} 1+\tanh(K_{j-\h}+K_{j+\h}), \nonumber \\
\uparrow \; \uparrow \; \downarrow \; \to  \; \uparrow \;
\downarrow \; \downarrow \mbox{\quad and\quad} \downarrow\;
\downarrow \; \uparrow \; \to \;\downarrow \; \uparrow \;
\uparrow  & \mbox{\quad with rate\quad} 1-\tanh(K_{j-\h}-K_{j+\h}),\nonumber \\
\downarrow \; \uparrow \; \uparrow \; \to \; \downarrow \;
\downarrow \; \uparrow \mbox{\quad and\quad} \uparrow \;
\downarrow \; \downarrow \; \to  \; \uparrow \; \uparrow \;
\downarrow  &\mbox{\quad with rate\quad}
1+\tanh(K_{j-\h}-K_{j+\h}),\qquad
\end{align}
where $\Gamma_j$'s have been taken independent of $j$, and set to
one by rescaling the time. As it could be expected, the left-right
symmetry is violated, but the up-down symmetry is not. The
particle reaction-diffusion picture turns into following
reaction-diffusion model, which is not left-right symmetric, as
expected.
\begin{align}\label{ihg.14}
\circ\, \circ \to \, \bullet\, \bullet &\mbox{\quad with rate\quad} 1-\tanh(K_{j-\h}+K_{j+\h}), \nonumber \\
\bullet\, \bullet \to \, \circ\, \circ &\mbox{\quad with rate\quad} 1+\tanh(K_{j-\h}+K_{j+\h}), \nonumber \\
\circ\, \bullet \to \, \bullet\, \circ &\mbox{\quad with
rate\quad} 1-\tanh(K_{j-\h}-K_{j+\h}), \nonumber \\
\bullet\, \circ \to \, \circ\, \bullet &\mbox{\quad with
rate\quad} 1+\tanh(K_{j-\h}-K_{j+\h}).
\end{align}

The evolution equation for the expectation values of the spins in
the bulk turns out to be
\begin{align}\label{ihg.15}
\langle\dot s_j\rangle=\;&-2\,\langle s_j\rangle+
[\tanh(K_{j-\h}+K_{j+\h})+\tanh(K_{j-\h}-K_{j+\h})]\,\langle
s_{j-1}\rangle\nonumber \\ & +
[\tanh(K_{j-\h}+K_{j+\h})-\tanh(K_{j-\h}-K_{j+\h})]\, \langle
s_{j+1}\rangle, \quad 0<j<L.
\end{align}
One has to add two other equations governing the evolution of
$s_0$ and $s_L$. These are of the form
\begin{align}\label{ihg.16}
\langle\dot s_0\rangle=\;&a_{-1}+a_0\,\langle
s_0\rangle+a_1\,\langle s_1\rangle,\nonumber\\
\langle\dot s_L\rangle=\;&a_{L+1}+a_L\,\langle
s_L\rangle+a_{L-1}\,\langle s_{L-1}\rangle,
\end{align}
where $a_j$'s are constants.
\section{The static solution}
For the static solution ($\langle s\rangle_{\st}$), the left hand
side of (\ref{ihg.15}) vanishes and one obtains
\begin{align}\label{ihg.17}
\langle s_{j+1}\rangle_\st=&\;
-\frac{\sinh(2\,K_{j-\h})}{\sinh(2\,K_{j+\h})}\,\langle
s_{j-1}\rangle_\st\nonumber\\
&\;+ \frac{\cosh(2\,K_{j-\h})+\cosh(2\,K_{j+\h})}
{\sinh(2\,K_{j+\h})}\,\langle s_j\rangle_\st, \quad 0<j<L,
\end{align}
which  can be written as following matrix form
\begin{equation}\label{ihg.18}
X_{j+\h}=D_j\,X_{j-\h},
\end{equation}
where
\begin{equation}\label{ihg.19}
X_\alpha:=\begin{pmatrix}\langle s_{\alpha-\h}\rangle_\st\\
\langle s_{\alpha+\h}\rangle_\st\end{pmatrix},
\end{equation}
and
\begin{equation}\label{ihg.20}
D_j:=\begin{pmatrix}0&1 \\ & \\
\displaystyle{-\frac{\sinh(2\,K_{j-\h})} {\sinh(2\,K_{j+h})}}&
\displaystyle{\frac{\cosh(2\,K_{j-\h})+\cosh(2\,K_{j+\h})}
{\sinh(2\,K_{j+\h})}}\end{pmatrix}.
\end{equation}
One can write $D_j$ as
\begin{equation}\label{ihg.21}
D_j:=\Sigma_{j+\h}\,\Delta_j\,\Sigma^{-1}_{j-\h},
\end{equation}
where
\begin{equation}\label{ihg.22}
\Sigma_\alpha:=\begin{pmatrix}\cosh K_\alpha &\sinh K_\alpha \\
\sinh K_\alpha &\cosh K_\alpha\end{pmatrix},
\end{equation}
and
\begin{equation}\label{ihg.23}
\Delta_j:=\begin{pmatrix}
\displaystyle{\frac{\sinh K_{j-\h}}{\cosh K_{j+\h}}}& 0 \\ & \\
0 & \displaystyle{\frac{\cosh K_{j-\h}}{\sinh
K_{j+\h}}}\end{pmatrix}.
\end{equation}
Using (\ref{ihg.18}), one arrives at
\begin{equation}\label{ihg.24}
X_\alpha=D_{\alpha,\beta}\,X_\beta,
\end{equation}
where
\begin{align}\label{ihg.25}
D_{\alpha,\beta}&:=\Sigma_\alpha\,\Delta_{\alpha,\beta}\,\Sigma^{-1}_\beta,\\
\label{ihg.26}
\Delta_{\alpha,\beta}&:=\Delta_{\alpha-\h}\cdots\,\Delta_{\beta+\h},
\end{align}
so that,
\begin{equation}\label{ihg.27}
\Delta_{\alpha,\beta}:=\begin{pmatrix}
\displaystyle{\Mu_{\alpha,\beta}\,\frac{\sinh K_\beta}{\cosh K_\alpha}}& 0 \\
& \\
0 & \displaystyle{\Mu_{\alpha,\beta}^{-1}\,\frac{\cosh
K_\beta}{\sinh K_\alpha}}\end{pmatrix},
\end{equation}
and
\begin{equation}\label{ihg.28}
D_{\alpha,\beta}=\begin{pmatrix}
\displaystyle{\left(\Mu_{\alpha,\beta}-\frac{1}{\Mu_{\alpha,\beta}}\right)\,\sinh
K_\beta\,\cosh K_\beta}&\displaystyle{\frac{\cosh^2 K_\beta}{
\Mu_{\alpha,\beta}}-\Mu_{\alpha,\beta}\,\sinh^2 K_\beta} \\ & \\
\displaystyle{\left(\Lambda_{\alpha,\beta}-\frac{1}{\Lambda_{\alpha,\beta}}\right)\,\sinh
K_\beta\,\cosh K_\beta}&\displaystyle{\frac{\cosh^2 K_\beta}{
\Lambda_{\alpha,\beta}}-\Lambda_{\alpha,\beta}\,\sinh^2 K_\beta}
\end{pmatrix},
\end{equation}
where
\begin{align}\label{ihg.29}
\Lambda_{\alpha,\beta}&:=\tanh K_\alpha\cdots\tanh K_{\beta+1},\\
\label{ihg.30} \Mu_{\alpha,\beta}&:=\tanh K_{\alpha-1}\cdots\tanh
K_{\beta+1}.
\end{align}
The boundary conditions (\ref{ihg.16}) are
\begin{align}\label{ihg.31}
A_\h\,X_\h&=-a_{-1},\nonumber\\
A_{L-\h}\,X_{L-\h}&=-a_{L+1},
\end{align}
where
\begin{equation}\label{ihg.32}
A_\alpha:=\begin{pmatrix}a_{\alpha-\mu}&a_{\alpha+\mu}\end{pmatrix}.
\end{equation}

The steady state profile near the end-site $0$ is determined by
the eigenvalues of the matrix $D_{\alpha,\h}$, where $\alpha$ is
some site far from the ends. One has
\begin{align}\label{ihg.33}
X_\alpha&=X_\alpha^a\,\fb_a,\nonumber\\
X_{\h}&=X_{\h}^a\,\fb_a,
\end{align}
where $\fb_a$ is the eigenvector of $D_{\alpha,\h}$ corresponding
to the eigenvalue $\lambda^a$, and $X_\alpha^a$'s and $X_{\h}^a$'s
are the coefficients of expansions of $X_\alpha$ and $X_{\h}$ in
terms of the eigenvectors. It is seen that
\begin{equation}\label{ihg.34}
X_{\alpha}^a=\lambda^a\,X_\h^a.
\end{equation}
While the exact values of these coefficients are determined by the
boundary conditions (\ref{ihg.31}), one can say whether in the
thermodynamic limit these coefficients vanish or not, without
referring to the exact from of the boundary conditions.

In the thermodynamic limit, corresponding to each eigenvalue, two
cases may occur.
\begin{itemize}
\item[\textbf{i}] The eigenvalue $\lambda^a$ tends to infinity.
In this case $X_{\h}^a$ tends to zero.
\item[\textbf{ii}] The eigenvalue $\lambda^a$ tends to zero or a
finite number. In this case $X_{\h}^a$ is generally nonzero.
\end{itemize}
Obviously, similar cases occur at the other end point. It is seen
that this behavior (some eigenvectors missing or not in the
solution corresponding to the end points) at one of the end points
is independent of the analog behavior at the other end.

\section{Some examples}
Consider some special cases.
\begin{itemize}
\item[\textbf{1}] Constant coupling:
\begin{equation}\label{ihg.35}
K_\alpha=K.
\end{equation}
In this case the relation of $D_{\alpha,\beta}$ with
$\Delta_{\alpha,\beta}$ is a similarity transformation. So the
eigenvalues $D_{\alpha,\beta}$ are the diagonal elements of
$\Delta_{\alpha,\beta}$, which are
\begin{align}\label{ihg.36}
\lambda^1&=\tanh^{\alpha-\h} K,\nonumber\\
\lambda^2&=\coth^{\alpha-\h} K,
\end{align}
showing that one of the eigenvalues is larger and the other is
smaller than one. So only one of the eigenvectors enters $X_\h$.
This is regardless of the value of $K$. So there is no static
phase transition, as it was seen in the case of ordinary
(symmetric) Glauber model, \cite{MA}.
\item[\textbf{2}] Periodic coupling:
\begin{equation}\label{ihg.37}
K_{\alpha+m}=K_\alpha.
\end{equation}
In this case the behavior of the eigenvalues of $D_{\alpha,\h}$ is
determined by the eigenvalues of $D_{\alpha+m,\alpha}$, which are
$\Lambda_{\alpha+m,\alpha}$ and $\Lambda^{-1}_{\alpha+m,\alpha}$.
This shows that one of the eigenvalues is greater than one and the
other is less than one. So the situation is similar to that of
constant coupling.
\item[\textbf{3}] Defects in the lattice:
No new phenomena is seen, as long as the defects are localized,
i.e. they are for from the end points. So if there is a lattice
that has some defects but otherwise is uniform, the static
behavior near the end points is similar to that of a uniform
lattice.
\item[\textbf{4}] A lattice with different behavior at different end points:
The behaviors of the static solution near the two ends are
independent of each other, provided the behavior change occurs far
from the end points. So all the phenomena seen in previous special
cases can be seen at each end point, independent of the other end
point.
\item[\textbf{5}] A lattice with different signs of coupling constants:
One can define the new variables $s'$ and couplings $K$ as
\begin{align}\label{ihg.38}
s'_j&:=[\sgn(K_\h)\cdots\sgn(K_{j-\h})]\,s_j,\\ \label{ihg.39}
K'_\alpha&:=|K_\alpha|
\end{align}
where $\sgn$ is the sign function, and investigate the system in
terms of these. So nothing new happens.
\item[\textbf{6}] A lattice with increasing coupling constants
(increasing towards an end point): Suppose that $K_\alpha$ is an
increasing function of $\alpha$, so that
\begin{align}\label{ihg.40}
\lim_{\alpha\to\infty}K_\alpha&=\infty,\\ \label{ihg.41}
\lim_{\alpha\to\infty}\Lambda_{\alpha,\beta}&=\Lambda,
\end{align}
where $\Lambda$ is neither zero nor infinity. Using
(\ref{ihg.29}), it is seen that (\ref{ihg.41}) implies
\begin{equation}\label{ihg.42}
\lim_{\alpha\to\infty}\tanh K_\alpha=1,
\end{equation}
which is equivalent to (\ref{ihg.40}). The criterion
(\ref{ihg.41}) itself, is equivalent to
\begin{equation}\label{ihg.43}
\lim_{\alpha\to\infty}[\exp(-2\,K_{\beta+1})+\cdots+\exp(-2\,K_\alpha)]=\ell,
\end{equation}
where $\ell$ is finite. Assuming that this is the case, it is seen
that $D_{\alpha,\beta}$ tends to a finite matrix $D$, where
\begin{equation}\label{ihg.44}
D=\begin{pmatrix}
\displaystyle{\left(\Lambda-\frac{1}{\Lambda}\right)\,\sinh
K_\beta\,\cosh K_\beta}&\displaystyle{\frac{\cosh^2
K_\beta}{\Lambda}-
\Lambda\,\sinh^2 K_\beta} \\ & \\
\displaystyle{\left(\Lambda-\frac{1}{\Lambda}\right)\,\sinh
K_\beta\,\cosh K_\beta}&\displaystyle{\frac{\cosh^2 K_\beta}{
\Lambda}-\Lambda\,\sinh^2 K_\beta}
\end{pmatrix}.
\end{equation}
So in this case both of the eigenvectors remain in $X_\h$.
\end{itemize}
The last case shows that there is a static phase transition. In
one phase only one eigenvector remains in $X_\h$, while in the
other phase both eigenvectors remain.

\section{Concluding remarks}
An Ising model with nonuniform coupling constants on a
one-dimensional lattice with boundaries was studied. Based on
detailed balance, the evolution of this model was investigated,
from which an equation for the static solution was obtained. While
it is known that ordinary one-dimensional Ising model does not
exhibit any phase transition at nonzero temperatures, here the
inhomogeneity at the boundary conditions does permit such a phase
transition, while the exact form of the boundary conditions do not
enter the static phase transition studied here. The static phase
picture of the system was studied in general, and some examples
were given including one exhibiting a static phase transition.\\
\\
\textbf{Acknowledgement}:  This work was partially supported by
the research council of the Alzahra University.

\end{document}